\documentclass[preprint]{revtex4}
\usepackage{natbib}
\usepackage{amsmath}
\usepackage{graphicx}
\usepackage{epsfig}

\begin{document}
\title{Eigenmode frequency distribution of rapidly rotating neutron stars}
\date{\today}
\author{Stratos Boutloukos}\altaffiliation[Current address: ]{Department of Physics, University of Illinois at Urbana-Champaign, 61801, IL, USA}\email{stratos@uiuc.edu}
\author{Hans-Peter Nollert}
\affiliation{Theoretical Astrophysics, University of T\"ubingen,
                Auf der Morgenstelle 10, 72076, Germany}
\begin{abstract}

We use perturbation theory and the relativistic Cowling approximation to 
numerically compute characteristic oscillation modes of rapidly rotating
relativistic stars which consist of a perfect fluid obeying a
polytropic equation of state. 
We present a code that allows the computation of modes of arbitrary order.
We focus here on the overall distribution of frequencies.
As expected, we find an infinite pressure mode spectrum extending to
infinite frequency. In addition we obtain an
infinite number of inertial mode solutions confined to a finite, well-defined
frequency range which depends on the compactness and the 
rotation frequency of the star. 
For non-axisymmetric modes we observe how this range is shifted with
respect to the axisymmetric ones, moving towards negative frequencies and
thus making all $m>2$ modes unstable.
We discuss whether our results indicate that the star's spectrum must
have a continuous part, as opposed to simply containing an infinite number
of discrete modes.

\end{abstract}
\maketitle

\section{Introduction}
\label{intro}
Helioseismology and asteroseismology have provided a wealth of
information about our sun, early type variable stars and even white dwarfs
\citep[see][chap.\ 2]{unno}.
However, for neutron stars, this field is still at its beginning, since it is
very difficult to measure their oscillation modes in the 
electromagnetic spectrum.
The gravitational wave spectrum, though, will likely allow us to study
neutron star oscillations and obtain information about their parameters,
such as mass, radius, rotation rate and equation of state \citep{aster}.
In this context, a discovery by \citet{nils} caused
some excitement: the so-called $r$-modes may be unstable at any non-zero
rotation rate of a compact star. Once excited --- e.g. during the birth
of a neutron star, via an accretion process, or 
through tidal interactions with a bound compact object --- 
it may grow to sufficient strength to be detected by
an earth-bound gravitational wave detector \citep{nils03}. 
This mechanism, the so-called CFS instability \citep{cfs}, may cause
the fundamental pressure mode ($f$-mode) of a neutron star to become unstable as
well. However, this occurs only when the rotation frequency of the star
is close to the Kepler limit \citep{neutral}.
Many gravity modes ($g$-modes) can become unstable as well, although their growth
is most likely suppressed by viscous dissipation \citep{lai}.
The latter are related to either composition gradient or finite temperature,
which are not considered in this work.

In this study we concentrate on the overall distribution of
eigenfrequencies rather than looking at individual modes. The
identification of individual modes will be the focus of a subsequenct
paper. Studying the spectrum as a whole is useful to shed light on
questions such as: What are the frequency ranges where the star may
oscillate at all? How do they depend on the physical parameters of the
star, such as its rotation rate? Which of them may become unstable? Does
the spectrum have a continuous part, and are there discrete modes within
the frequency range of a continuous spectrum?

We concentrate here on isentropic models, where a star's
equilibrium as well as its perturbations can be described with the same
one-parameter equation of state $p=p(\epsilon)$ and constant entropy \citep{lf99}
(also called barotropic \citep{law}).
Deviations from an isentropic model become important only if the angular
spin frequency 
is comparable to or smaller than the Brunt-V\"ais\"al\"a frequency, which for 
neutron stars is of the order of 100Hz \citep{laf}.

$r$-modes have originally been defined for Newtonian models as inertial
modes of axial type. For relativistic models this definition is
not as straightforward any more since the oscillation modes (except dipole modes) 
consist of a mixture of polar and axial contributions,
leading to what was called `hybrid` modes \citep{lf99}. 
Still, even in the relativistic case with rotation, it is possible to
assign an overall parity to a mode by looking at the dominant term in an
infinite series representation over the harmonic index $\ell$.
\citet{lf99} call these modes axial-led or polar-led.
If the spherical harmonic with lowest $\ell$ that contributes to the
velocity field is axial, then
the mode was refered to as 'generalized $r$-mode' \citep{lindblom}.
Following \citep{definition}, this terminology has been made
obsolete by a better understanding of the problem. In the following we
will therefore always use the term ``inertial modes''.

In the relativistic slow rotation approximation \citep{laf,lfa}, 
their frequencies are found to be up to 15\% lower than those of
their Newtonian counterparts.  Using the relativistic 'traditional'
approximation a similar shift was found~\citep{mina}, but it applies
to the frequencies in the corotating frame, rather than those in the
asymptotic inertial frame. In all cases, the shift is smaller for
higher order modes.

In non-rotating stars, inertial modes are degenerate at zero frequency
due to the absence of their restoring force, the Coriolis force.
Rotation breaks this degeneracy, and for isentropic stars
there is an infinite number of inertial modes
confined to a finite frequency range \citep{greenspan}. 
The range of frequencies they cover has been shown for Newtonian 
incompressible stars to extend from $(-2-m) \nu$ to $(2-m)\nu$, 
where $\nu$ is the rotational frequency of the star and $m$ the 
azimuthal index \citep{lindblom}. 
Later \citet{brink} computed a large set of such modes in the same framework
and confirmed 
modes up to 30th order to have frequencies confined within this range.
\citet{rsk03} studied the inertial mode spectrum for relativistic barotropic
(as well as non-barotropic) stars in the slow-rotation approximation
by including coupling of modes up to a maximum harmonic index $\ell_{max}$.
Next to distinct inertial modes they found a continuous spectrum 
whose width depends on the compactness of the star and on $\ell_{max}$. 
They all lie roughly between $-\nu_K$ and $\nu_K$ for $m=0$ and between $-2\nu_K$ and $0$ for $m=2$, where $\nu_K\approx 1.915\nu$ is the breakup frequency of their model.
For $\ell_{max}\rightarrow\infty$ the authors expect the continuous spectrum 
to fully cover this range, with individual modes still existing inside the
continuous spectrum. On the other hand,
\citet{law} argue in their appendix that only individuals modes should be
present. 
\citet{framedrag} suggested that the width of the continuous spectrum
depends on the range of values that the difference between the rotation
rate of the star $\Omega$ and the frame dragging $\omega$ can take.

All stars show pressure driven modes ($p$-modes). The fundamental $p$-mode,
having the lowest frequency, is called the $f$-mode. They were studied first
in non-rotating models.
In Newtonian gravity the perturbation equation describing their frequencies
is of Sturm-Liouville type \citep[see eg.][]{cowl},
the solution is an infinite set of modes with frequency range unbound from
above.
General relativity and rotation do not change this general picture, and
individual frequencies are only slightly
affected for a spin frequency up to about half the break-up frequency of
the star\citep{nik}.
The $f$-mode typically has a frequency around 2kHz, the lowest order $p$-mode
a somewhat higher one.
For a review regarding fluid oscillations of neutron stars see
\citep{hp} and \citep{kostas}, \cite{Nik:Rev} for rotating neutron stars, and
\cite{Nils+Kostas} for $r$-modes.

Many studies of neutron star oscillations use the so-called Cowling approximation \citep{cowl},
which in its relativistic version allows to work only with fluid displacements and 
neglect metric perturbations \citep{rel-cowl}.
Although $p$-modes can be computed quite accurately in this way \citep{shijun},
\citet{finn} questioned whether neglecting all the metric perturbations is a good approximation
for modes involving large fluid velocities, such as the $g$-modes.
\citet{laf} extended this argument to the $r$-modes, which have similar properties,
and disputed the results of \citet{rsk03} due to this. 
\citet{yosh2e} compared the frequencies of the fundamental $r$-mode
for slowly rotating relativistic stars with and without the relativistic Cowling approximation 
and found the difference to be only a few percent of the rotational frequency of the star.
The same comparison for rapid rotation is not possible due to the lack of results with the full
spacetime perturbations, but for Newtonian stars the relative difference in the frequency is a few 
percent \citep{yosh2e}.
The last two results indicate that even for rapidly rotating relativistic stars
the Cowling approximation would give a good estimate of the $r$-mode frequencies.

On the other hand, using the slow-rotation approximation may be more
problematic than the Cowling approximation. It seems that going beyond the
first order in the rotation rate of the star, results may differ
considerably from those obtained in first order \citep{slow-rot}.
Therefore, in this paper we will discard the slow-rotation limit,
performing mode calculations for rapidly rotating stars using linear
perturbation theory in the Cowling approximation in relativistic
rather than Newtonian gravity. Given the results discussed above, 
we expect to find an infinite number of pressure modes with frequencies
on the order of kHz and higher, and possibly an infinite number
of inertial modes with frequencies dependent on the rotation rate of the star.

In section \ref{sec:2} we formulate the problem and discuss the
numerical setup. In section \ref{sec:axisym} we investigate axisymmetric
perturbations, first in the non-rotating case for comparison with known
results, and then for the rotating case. Results for non-axisymmetric
perturbations and a comparison with the axisymmetric ones
follow in section \ref{mneq0}, concluding with a discussion of the
results in Section \ref{concl}.

\section{Problem set-up}
\label{sec:2}
\subsection{Equilibrium background}

Neutron stars have been observed to spin up to millisecond periods. This
results in significant deviations from spherical symmetry both for the
fluid configuration of the star and for the spacetime.
Equilibrium solutions will thus not be spherically symmetric, but axisymmetric
at most. 
The general form of an axisymmetric metric describing a rotating body 
can be written as:
\begin{eqnarray}{\label{metric}}
ds^2 = -e^{\gamma+ \rho} dt^2 
      + e^{2 \alpha} \left(dr^2 + r^2 d \theta^2 \right)
      + e^{\gamma - \rho} r^2 \sin^2 \theta 
                        \left( d \phi - \omega dt \right)^2.
\end{eqnarray} 
where $r$,$\theta$,$\phi$ are quasi-isotropic coordinates, reducing to isotropic
ones if the rotation rate goes to zero\citep[see][]{Nik:Rev}.
Restricting ourselves to uniform rotation with frequency $\Omega$, 
the 4-velocity of the fluid is given by 
$u^{\alpha}=U^0\{1,0,0,\Omega\}$, with an energy-moment tensor
 $T_{\mu\nu}  = p\,g_{\mu\nu}+ (p+ \epsilon)\,U_\mu U_\nu$, assuming
that the star consists of an ideal fluid. 
For the equation of state we use a simple polytropic model of the form
$p=k\times \rho_0^{\gamma}$. Using a realistic equation of state
in tabulated form would not affect our analysis or our numerical procedure. 
In this paper, we restrict ourselves to a polytropic equation of state to facilitate
comparison of our results with previous studies by other authors.

We use the RNS code of \citet{rns} for computing the background models.
The same code was also used by \citet{nik}; we will occasionally refer
to their results for comparison.
Table \ref{tab:models} summarizes some of their models which
we will be using here. For all models, the parameters in the equation of
state are $k=217.856\text{km}^2$ and $\gamma = 2$; 
the central density is always
$\epsilon_c=0.894\times 10^{15}\text{g} / \text{cm}^3$.

\begin{table}[htb]
    \caption{The models of \citet{nik} which we refer to in this
      paper. Next to the resulting mass and radius we show both the angular frequency, 
      that is used in our analysis, and the natural rotation rate.}
  \begin{center}
    \begin{ruledtabular}
    \begin{tabular}[t]{rrrrr}
Model  &  Gravitational Mass    &  Radius  & $\Omega$ & Rot. rate ($\nu$)\\
BU0    & $ 1.4 \text{M}_{\odot}$   & 14.15 km  &  0         & 0\\
BU1    & $ 1.432 \text{M}_{\odot}$ & 14.51 km  &  2.185 kHz & 348Hz\\
BU6    & $ 1.627 \text{M}_{\odot}$ & 17.25 km  &  4.984 kHz & 793Hz\\
    \end{tabular}
    \end{ruledtabular}
    \label{tab:models}
  \end{center}
\end{table}

\subsection{Perturbations}

Assuming small deviations for the fluid variables we study linearized
perturbations on these stationary configurations. Since the background is
not spherically symmetric it is not helpful to decompose the perturbations
into spherical harmonics. Instead we exploit the axisymmetry of the
background by writing the perturbation quantities as
\begin{subequations}
  \label{eq:perturb}
\begin{eqnarray}
  \delta p &=& H e^{im\phi}
\end{eqnarray}
\begin{eqnarray}
  \delta u_{\alpha} &=& \frac{1}{p+\epsilon}
                        \{-\Omega f_3,f_1,f_2,f_3\}e^{im\phi},
\end{eqnarray}
\end{subequations}
where $H$,$f_i$ are functions of $t$, $r$ and $\theta$.
The reason for this particular choise will become clear in the next subection
(\ref{sec:numerics}).
The perturbation of the energy-density is related to the pressure
perturbation through 
\[ \delta p = \frac{d p}{d\epsilon}\delta\epsilon
            \equiv C_s^2\delta\epsilon  ,
\]
where $C_s$ is the speed of sound.

As mentioned earlier we use the Cowling approximation, 
(\citep{rel-cowl} after \citep{cowl,emden})
neglecting the metric perturbations.
This will not allow us to calculate any damping of the modes due to
emission of gravitational waves, but we can estimate the oscillation
frequencies and study the overall structure of the spectrum (see also section
\ref{intro}). 

The equations that describe the behavior of the perturbed quantities
arise from the perturbed form of the conservation of energy-moment
(equations of motion for the fluid):
\begin{eqnarray}{\label{eom}}
\delta\left(T^{\nu}_{\mu;\nu}\right) 
  = g^{\kappa \nu}(\delta T_{\mu\kappa,\nu}
  - \delta T_{\mu\rho}\Gamma_{\nu\kappa}^{\rho} 
  - \Gamma_{\mu\nu}^{\rho}\delta T_{\rho\kappa}) = 0
\end{eqnarray}
In general, these yield four independent partial differential equations
which are of first order in time and space.
Time-evolving sets of lengthy equations can often lead to instabilities
\citep[see eg.\ ][]{adam}.
We perform mode calculations by assuming a harmonic time-dependence
$e^{i\sigma t}$ for all four variables 
(e.g. $H (t,r,\theta)= \tilde{H} (r,\theta) e^{i\sigma t}$),
searching for frequencies
which allow non-trivial solutions of the perturbation equations.

\subsection{Numerical procedure\label{sec:numerics}}

Here we briefly describe the numerical method used in this work.
We discretize the system of equations at every grid point, including the
boundaries by making use of the boundary conditions. This results in a
system of linear equations of the form
\begin{equation}\label{bvp}
  {\mathbf A} \cdot X = 0
\end{equation}
where the unknowns \( X \) are the discrete perturbation quantities at
all grid points, and the coefficient matrix \( \mathbf A \) represents
the equations resulting from the perturbation equations at all grid points and
from the boundary conditions. \( \mathbf A \) is a highly sparse
matrix. The components of \( \mathbf A \) depend on the frequency
parameter $\sigma$.  It is now possible to examine the condition of 
\( \mathbf A \), searching for values of $\sigma$ which make 
\( \mathbf A \) degenerate, since this is the only way that
Eq.\ (\ref{bvp}) allows non-trivial solutions for the perturbation
quantities. 

In the case we study here, we may rewrite Eq.\ (\ref{bvp}) as an eigenvalue
problem of the form
\begin{equation}\label{eigen}
  \mathbf{\tilde A} \cdot X = i\sigma X
\end{equation}

Note that this is a special case and not generally possible; it will probably
not work for the general perturbation equations which result if one does not
use the Cowling approximation.

We may now use a standard routine for finding eigenvalues of
Eq.\ (\ref{eigen}). We use the routine \texttt{cg.f} from the EISPACK
package of the NETLIB libraries which handles general complex matrices
by use of the QR-algorithm.  The characteristic frequencies of the star's
oscillation modes are obtained as eigenvalues of Eq.\ (\ref{eigen}), the
perturbation quantities \( X \) are the corresponding eigenfunctions.
It has turned out that a two-sided differencing scheme does not give
consistent results (see \citep{neb11}). We therefore use a one-sided
differencing scheme for all first derivatives throughout this work.

We need to take into account boundary conditions at the center and at
the surface of the star, and a regularity condition in the angular
direction. At the center, all variables are required to vanish by the
regularity condition there. We implement this condition simply by
setting all variables to zero at $r=0$. At the surface, all
perturbation variables must vanish: this follows from the definition of the
variables given in Eqs.\ (\ref{eq:perturb}),
since $p=\epsilon=\partial_r p=0$ (for polytropic equations of state)
and $\delta p\sim\partial_r p$ (to first order in a Taylor expansion).
For the special case of $m=0$,
the pressure perturbation is not required to vanish at the center;
its value must then be determined directly through Eqs.\ (\ref{eom}). 

For the boundaries in the $\theta$-direction one may use the fact that 
the rotational axis itself is special: For $m > 0$, all variables have to
vanish on the rotational axis due to regularity conditions 
\footnote{The $e^{im\phi}$-dependency of non-axisymmetric fluid perturbations requires
that on the axis all possible azimuthal angles give the same value, i.e. zero.}.
This can be used
directly to set values for the discretized variables there. For $m=0$ we may
construct a grid in such a way that no grid points fall on the rotational
axis. In this case, one must use the symmmetry condition to construct a
boundary condition in the $\theta$-direction. This technique is also
applicable for $m > 0$, keeping in mind that the symmetry is different for
even and odd values of $m$. We have used both techniques for the $m=2$ case
and found no significant difference. In order to make the code applicable
both in the axisymmetric and the non-axisymmetric case, we implemented
the second possibility.

The RNS code itself has finite resolution and thus introduces some
numerical error into the results. We will always see a combination of numerical
error coming from the RNS code and from our own code. When we attempt to
study convergence properties we therefore use a high but fixed
resolution of the RNS code, varying only the resolution of our eigenvalue
code. In order to avoid the use of interpolation and the additional error
associated with it, we run our code only with grid spacings which are
multiples of the fixed grid spacing we have chosen for the RNS code.


\section{Axisymmetric perturbations}{\label{sec:axisym}}

We first study axisymmetric perturbations as a test for our code. We use the
corresponding  equations in their general form, including all terms with
arbitrary $m$. Setting $m=0$ we select perturbations that are symmetric around
the rotation axis, including all $\ell \ge 0$ contributions.
Axisymmetric perturbations have been studied by \citet{nik} using nonlinear 
time-evolution; we will turn to their results for comparison. 
In order to gain a better understanding of the underlying physics and
numerics, we first solve Eqs.\ (\ref{eom}) for the non-rotating case where
expressions may be reduced considerably.
We then proceed to include rotation at arbitrary rotation rate.

\subsection{No rotation}
\label{no-rot}

Setting $\Omega=0$ in Eqs.\ (\ref{eom}) and using harmonic time-dependence
the system of equations for the variables defined in Eqs.\ \ref{eq:perturb}
takes the form:
\begin{eqnarray}{\label{nonrot}}
i\sigma H &=& -\frac{imC_s^2e^{2\rho}U^0}{r^2\sin^2{\theta}}f_3 -
               \frac{C_s^2}{e^{2\alpha}U^0} \left\{\frac{1}{r^2}
                 \left(\frac{\partial f_2}{\partial\theta} + 
                   \frac{\cos{\theta}}{\sin{\theta}}f_2\right)
             + \frac{\partial f_1}{\partial r} 
              + \left( \frac{3}{2}\frac{\partial\gamma}{\partial r} 
                + \frac{1}{2}\frac{\partial\rho}{\partial r}
                +\frac{2}{r} \right)f_1\right\}\nonumber\\
i\sigma f_3 &=& -\frac{im}{U^0} H, \qquad
i\sigma f_1 = -\frac{1}{U^0}\frac{\partial H}{\partial r}, \qquad
i\sigma f_2 = -\frac{1}{U^0}\frac{\partial H}{\partial \theta}
\end{eqnarray}
In this simple case, one variable ($H$) would be sufficient to describe the whole
oscillation problem since one can transform the system of
Eqs.\ (\ref{nonrot}) into a single second order equation.
Solving the latter gave the same results as solving Eqs.\ (\ref{eom}) for
$\Omega=0$, so we will always present results from the full first-order system 
for consistency.

Table \ref{bu0} shows the lowest resulting eigenfrequencies for the
model BU0 with increasing resolution in the radial direction, starting from
21 points. Increasing the resolution in the $\theta$ direction does not seem to
affect the results, and it is computationally expensive. We therefore
keep the $\theta$-resolution fixed for the results shown in
Table \ref{bu0}. From now on we will always use the physical
oscillation mode frequency $f$ rather than the angular frequency
($\sigma=2\pi f$), unless stated otherwise.

\begin{table}[hbt]
\caption{The frequencies ($\frac{\sigma}{2\pi}$) in Hz of the first three
  fundamental modes ($f$) and the lowest pressure mode ($p^1$) for
  axisymmetric oscillations of BU0, for several resolutions in the radial
  direction and a fixed number of points in the angular direction. In the
  next to last row the extrapolated values for 'infinite' radial
  resolution (see figure \ref{conv}) are listed, while the last row shows
  the results of \citet{nik}.
}\label{bu0}
\begin{center}
\begin{ruledtabular}
\begin{tabular}{llllll}
$n_{\theta}$& $n_r$ &  $f_{\ell=1}$ &  $f_{\ell=2}$ &  $f_{\ell=0}$ & $p^1_{\ell=2}$ \\
10  &21  & 1377 & 1974 & 3076 & 4627 \\
    &26 & 1367 & 1943  & 3032 & 4551 \\
    &51  & 1344 & 1872 & 2917 & 4344 \\
    &101 & 1328 & 1830 & 2840 & 4197 \\
\cline{2-6}
&$\infty$  & \bf{1317}$\pm2$ & \bf{1794}$\pm3$ & \bf{2787}$\pm11$ & \bf{4102}$\pm26$\\
\multicolumn{2}{l}{\citet{nik}}  & 1335 & 1846 & 2706 & 4100\\
\end{tabular}
\end{ruledtabular}
\end{center}
\end{table}

\begin{figure}[tbh]

\includegraphics[width=0.5\textwidth]{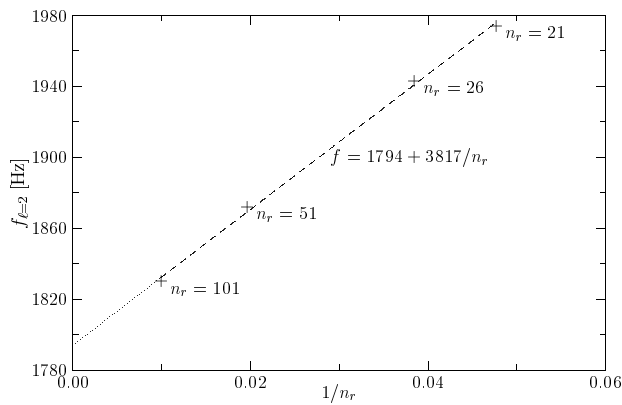}
\caption{\label{conv}The computed eigenfrequencies (here the $\ell=2$
  $f$-mode for BU0) follow an inverse power law. 
 We may extrapolate to $n_r \rightarrow \infty$ to obtain
  the frequencies for nominally infinite radial resolution.}
\end{figure}

Figure \ref{conv} shows a plot of the $(l=2,m=0)$ $f$-mode frequencies 
from Table \ref{bu0} 
as a function of the radial resolution. They follow an inverse power law
of the form \(f = f_\infty + \delta f / n_r + \dots \), exhibiting first order
convergence as one expects for one-sided differences. We may thus
extrapolate the computed values to obtain an eigenfrequency $f_\infty$
at nominally infinite resolution.  In general, the relative mean deviation of
the fitted numbers is less than 1\%. Note that there is an
additional systematic error of the same general magnitude, resulting from the
finite accuracy of the background quantities.

Since the background is now spherically symmetric, we can assign a
definite value of $\ell$ to the oscillation modes. Comparing the
extrapolated frequencies in Table \ref{bu0} with results by \citet{nik},
we conclude that they correspond to the 
fundamental modes that correspond to $\ell=0,1,2$ and the first
pressure mode for $\ell=2$,
with an agreement up to a few percent. 
The remaining differences are likely due to the finite resolution of the
background calculation.

The method remains basically the same when we turn on  rotation. We thus
expect to obtain the fluid modes with similar accuracy even for rapid
rotation. 
The picture may change, though, for the rotation-driven modes:
These are degenerate at zero frequency in the non-rotating case. Therefore,
there are no results for the non-rotating case that can be used as an
indication for the accuracy of our computational results.


\subsection{Rapid rotation}
\label{raprot}

With rotation being included the equations of motion for the fluid
become quite lengthy. Schematically they take the form
\begin{eqnarray*}
\Pi\frac{\partial H}{\partial t}&=&\mathcal{L}_{1}\left(H,f_3,f_1,f_2,\frac{\partial f_1}{\partial r},\frac{\partial f_2}{\partial\theta},-AC_s^2\frac{\partial f_3}{\partial t}\right)\\
ZC_s^2\frac{\partial f_3}{\partial t}&=&\mathcal{L}_{2}\left(H,f_3,f_1,f_2,\frac{\partial f_1}{\partial r},\frac{\partial f_2}{\partial\theta},-B\frac{\partial H}{\partial t}\right)\\
\frac{\partial f_1}{\partial t}&=&\mathcal{L}_{3}\left(f_3,f_1,\frac{\partial H}{\partial r}\right)\\
\frac{\partial f_2}{\partial t}&=&\mathcal{L}_{4}\left(f_3,f_2,\frac{\partial H}{\partial\theta}\right),
\end{eqnarray*}
where A,B,Z,$\Pi$ are functions of the background quantities 
(provided in Appendix \ref{eqs}).
However, the frequency $\sigma$ now appears in the matrix $\mathcal{L}$ on
the right hand side. Therefore, this form is not suitable for 
performing an explicit eigenvalue calculation as described
in Subsection \ref{sec:numerics}. 
We thus define new variables:
\begin{eqnarray*}
F&=&\Pi H+A C_s^2 f_3\\
V&=&\text{Z}C_s^2f_3+B H ,
\end{eqnarray*}

which together with $f_1,f_2$ form the set of variables we will use from now on.
The full set of equations is provided in Appendix \ref{eqs}.

\begin{figure}[tb]
\begin{center}
\includegraphics[width=0.5\textwidth]{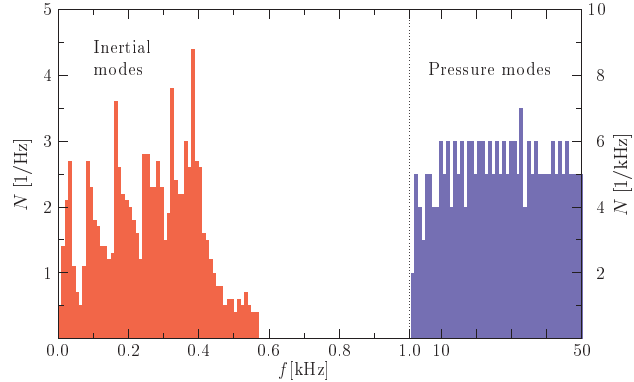}
\end{center}

\caption{\label{hist}A histogram of the number (per frequency bin) of
  all solutions returned by the eigenvalue code for the $m=0$ eigenvalue
  problem of the model BU1. The frequency bins have a constant width
  of 10 Hz for the inertial mode frequencies and 1 kHz for the
  pressure mode frequencies. Note the change in frequency scale at 1kHz.}
\end{figure}

\begin{table}[b]
\caption{Frequencies (in kHz) of two fundamental
$m=0$ pressure-driven oscillations for the polytropic models BU1 and BU6,
compared with the results of \citep{nik}.
In the non-rotating limit these correspond to $l=0$ and $l=2$ modes.
Again, our values have been extrapolated to nominally infinite radial
resolution.
Convergence in this case was not as clean as shown in Fig.\ \ref{conv},
especially for the rapidly rotating model BU6. This is reflected in larger 
uncertainty estimates.
\label{tab:axisym}}
\begin{center}
\begin{ruledtabular}
\begin{tabular}{lllll}
& $f_{l=0}$ (BU1) & $f_{l=2}$(BU1) & $f_{l=0}$(BU6) & $f_{l=2}$(BU6)\\
This paper & 2.720$\pm20$ & 1.834$\pm25$ & 2.292$\pm193$ & 1.718$\pm85$\\
\citet{nik} & 2.657 & 1.855 & 2.456 & 1.762\\
\end{tabular}
\end{ruledtabular}
\end{center}
\end{table}

In Fig.\ \ref{hist} we show all solutions returned by the eigenvalue code
for axisymmetric perturbations of model BU1, using a low radial
resolution of $n_r =25 $.  As far as results for pressure modes are
available from the literature we list them along with our results in
Table \ref{tab:axisym}. 
No previous results are available for axisymmetric inertial modes of rapidly
rotating relativistic stars; see \citet{rsk03} for modes of slowly rotating
stars. 
Very recently \citet{nikCFC} published frequencies of the three strongest
axisymmetric inertial modes in the conformally flat approximation. 
They all lie inside the corresponding inertial mode spectrum (see below).

We notice from Fig.\ \ref{hist}
that the eigenvalues are clustered into two groups, one above 1000Hz
and one below. These correspond to the expected frequency ranges for
pressure modes and inertial modes. The latter range is more densely
populated; for example, we see more solutions between, say, 500 and 520 Hz
than between 2 and 5 kHz. This is not as surprising as it may seem at
first: according to theory (see eg.\ \citep{law} for an overview table),
there is an infinite number of pressure modes, with frequencies
extending to infinity. For the inertial modes, though, one expects an
infinite number of modes as well, but confined to a well defined
frequency range. According to our computation this range appears to extend
from 0 
to about 600 Hz for the model BU1.

\begin{figure}[t]
\begin{minipage}[t]{0.48\textwidth}
\includegraphics[width=\textwidth]{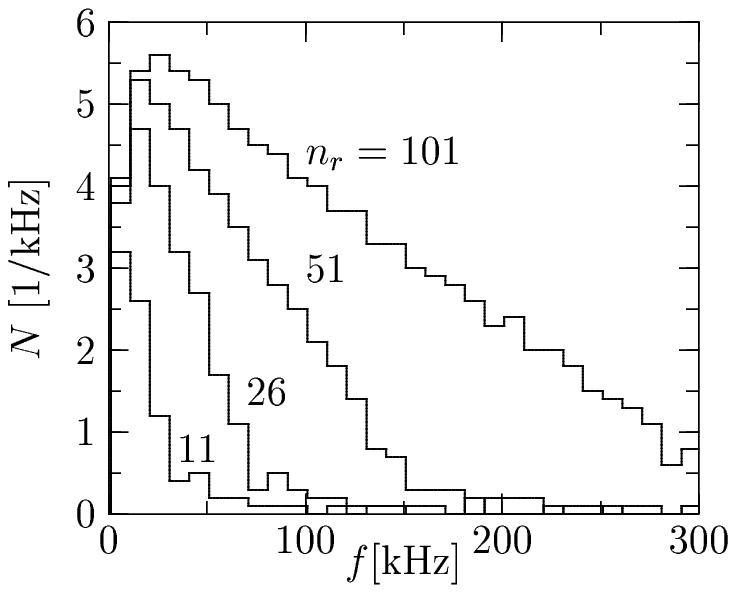}
\caption{\label{hist_f}A histogram of the number of p-modes per frequency bin computed for 
the $m=0$ eigenvalue problem of the model BU1. The bins are equally
sized at 10kHz}
\end{minipage}
\hfill
\begin{minipage}[t]{0.48\textwidth}
\includegraphics[width=\textwidth]{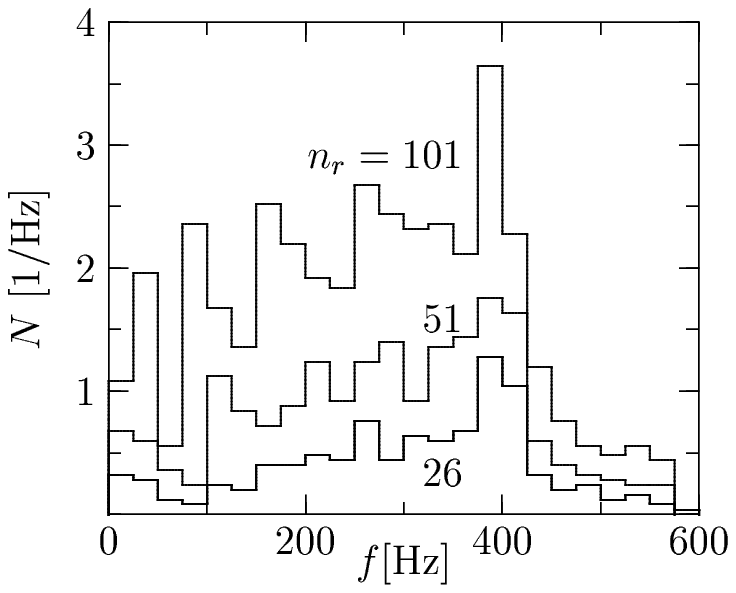}
\caption{\label{hist_i} Same as Fig.~\ref{hist_f}, but for the inertial
  modes of the same problem. The bin size is now 25Hz}
\end{minipage}
\end{figure}

In Figs.\ \ref{hist_f} and\ \ref{hist_i} we show the $p-$mode and
inertial mode ranges separately for increasing radial resolution. In
both ranges the number of frequency eigenvalues increases with
increasing resolution. There is an important distinction, however: in
the $p$-mode range, higher frequency ranges are increasingly populated
as the resolution increases, while the population of low frequencies
approaches a limiting value. We observe that as the radial resolution $n_r$
increases, the number of computed eigenvalues increases as $10\times
n_r$. The lowest $p$-mode frequency, which belongs to the $m=2$ $f$-mode,
is roughly the same for all resolutions. It shifts
in frequency by a few hundred Hz, which is just a few percent of the bin
size.  Typically the number of solutions per frequency bin approaches a limit
value $\alpha$ like $N_{\rm bin} \approx \alpha + \beta / n_r$ with some
constant $\beta$.

For each stellar model 
the maximum value of frequency eigenvalues tends to
infinity for increasing resolution. This is similar to the situation
for quasi-normal modes of black holes, with an infinite number of modes
which is not confined to a finite part of the complex frequency plane
(see \citep{nollert}). 

In the inertial mode range, on the other hand, the frequency range does not
change as the 
resolution increases; in fact, the upper limit (600Hz for model BU1) is quite
robust.
Instead, the population increases fairly homogeneously over
the whole frequency range.  The total number of solutions calculated
develops, just as for the p-modes, like $10\times n_r$. The number of
points per bin grows linearly; for the bin e.g. around 400Hz, as
$N_{\rm bin}\simeq 2\times n_r$. This is a strong indication that an
infinite number of solutions exists in this frequency range. It
would seem likely that there is an infinite number of physical modes in
this range as well.

\begin{figure}[htb]
\begin{center}
\includegraphics[width=0.5\textwidth]{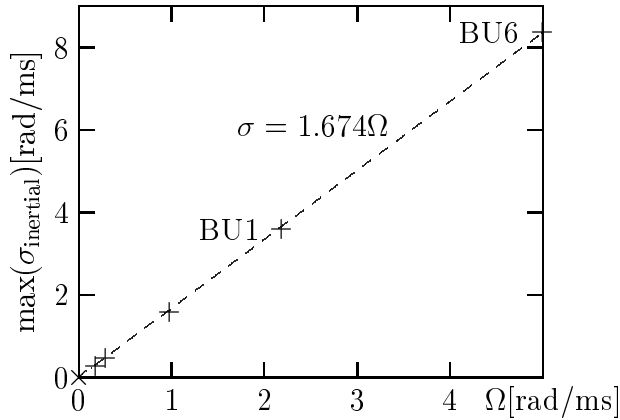}
\end{center}
\caption{\label{fig:fmax}The highest inertial mode angular
  frequency for $m = 0$ as a function of the star's rotational angular
  frequency, starting at the model BU0 (no rotation) and moving up to
  model BU1 (2185Hz).}
\end{figure}

While the upper limit for the frequency range does not depend on the
resolution used in the numerical calculation, it actually depends
linearly on the rotational frequency of the star, as shown in
Fig.\ \ref{fig:fmax}. The linear fit reveals 
$\sigma_{max}=1.674\times\Omega$ (where $\sigma$ and $\Omega$ are now the
angular frequencies), with a negligible statistical error.

The picture in Fig.\ \ref{hist_f} is actually quite common for numerical
studies of oscillation spectra:
As one increases the resolution, frequency ranges at increasingly higher
frequency become populated, while there is a finite limit for lower
frequency ranges. Usually one expects that at low resolution, only modes
with low frequencies can be computed reliably, since they are the only
ones which can be resolved sufficiently well. Numerical solutions at
high frequencies are likely spurious and cannot be trusted. With
increasing resolution, the number of reliable solutions increases and
their range extends to higher frequencies. 

On the other hand, if we are confronted with a finite frequency range
which shows an increasing number of solutions with increasing
resolution, such a distinction cannot be made in a meaningful way. It is
therefore not clear which of the numerical solutions in the frequency
range corresponding to inertial modes should be considered physical
solutions, and which should be discarded as numerical artefacts.

Just as in the non-rotating case, one can establish convergence for any finite
value of the frequency in the upper range (Fig.\ \ref{hist_f}), since there is
a finite number of distinct modes in any finite frequency interval. The limit
values very likely correspond to physical modes of the star. In the lower
frequency range (Fig.\ \ref{hist_i}) it is not even clear how to establish a
correspondence between eigenvalues at increasing resolutions, let alone
define convergence and establish a correspondence to physical modes.

However, the number of eigenvalues increases linearly with the number                 
of grid points used in the calculation. We therefore find it unlikely                 
that this part of the spectrum contains a finite number of discrete                   
oscillation frequencies. If an infinite numbers of eigenvalues is                     
confined to a finite frequency range, then they must have at least one                
accumulation point. It is possible that there is only a finite number                 
of accumulation points, even only one. An example for such a system is                
an atom which has an infinite number of bound states confined to a                    
finite energy interval, with zero energy as the only accumulation                     
point. However, as we increase the resolution of our calculations, the                
number of eigenvalues increases rather evenly throughout the whole                    
frequency range in question, and there is no sign that there are parts                    
where the increase might saturate at a finite level. We thus consider                 
it likely that the oscillation frequencies form a dense set throughout                
the inertial mode frequency range. The spectrum may also be continuous                
over this whole range or in parts of it, but this can neither be                      
confirmed nor excluded based on the results of our code. 

\section{Non-axisymmetric perturbations}
\label{mneq0}

For $m>0$ the picture is more complicated but also more interesting than
the axisymmetric case since
the $m=2$ modes are the ones most unstable to gravitational radiation. 
Solving this system for $m=2$ and the BU1 model in the way described in
the previous section we find a set of eigenvalues containing both
positive and negative frequencies. In the axisymmetric case negative
eigenvalues are equivalent to the positive frequency solutions. 
Breaking axial symmetry changes frequencies (for an
asymptotic observer) towards negative values, so that much or even all of the
frequency range corresponding to inertial modes becomes negative.

In Newtonian gravity (see e.g.\ \citet{unno}), a polar mode of order 
$n$, harmonic index $\ell$, and frequency $\sigma_0$,
splits under rotation to $2\ell+1$ modes with 
frequencies changing to $\sigma = \sigma_0 - m\Omega E_{n \ell}\;(+O(\Omega^2))$
where $E_{n \ell}$ is a function depending on the eigenfunction 
of each mode. For low order pressure modes the value of this
function is about 0.1, so one would need rotational frequencies close to the
Kepler limit for the frequency to change sign.

However, there are modes which have $\sigma<0$ for \textit{any} rotation
rate, such as the $r$-modes, which, 
in the Newtonian, slow-rotation limit, have 
$\sigma=-2m\Omega\frac{(\ell-1)(\ell+2)}{\ell(\ell+1)}$.
This is the picture we find for $m=2$, with
all inertial modes having negative frequencies. Such a change of sign is
often used to mark the onset of the instability for the corresponding mode
\citep[see eg.\ ][]{Nik:Rev}.

In Fig.\ \ref{fig:fmax} we saw that the upper cutoff frequency for the
inertial mode range grows linearly with the rotational frequency of the
star. In the non-axisymmetric case, there is also a dependence on
$m$. For the series of models listed in Table \ref{tab:models} we obtain
a least square fit of $\sigma_{\rm max} = \Omega(1.6 - 1.05m)$, with
statistical errors of about $0.1$ for both values.
For less relativistic models with a central energy density 1/10 that of
BU1, this changes to $\sigma_{\rm max} = \Omega (1.93-1.03m)$, close to
what \citet{lindblom} predicted and \citet{brink} found for Newtonian stars
where $\sigma_{\rm max} \approx \Omega (2-m)$.

\begin{figure}[tb]
\vspace{2.\baselineskip}
\begin{minipage}{0.48\textwidth}
\includegraphics[width=\textwidth]{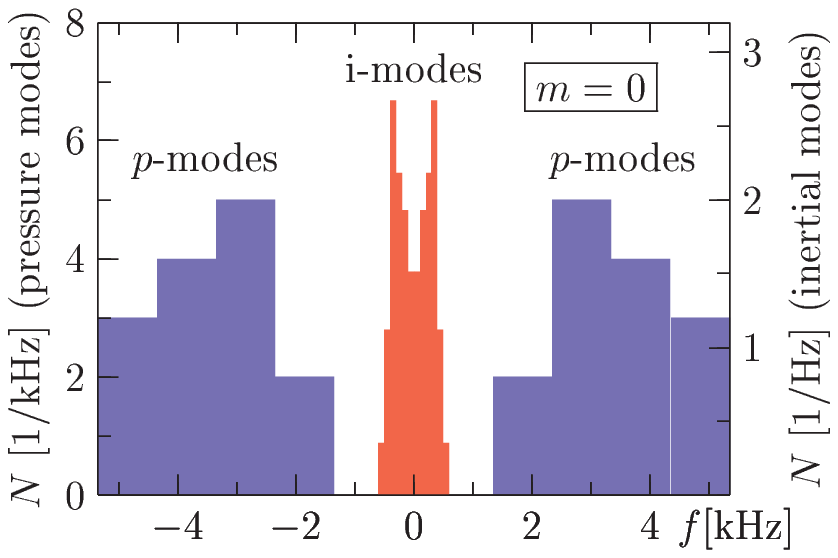}
\end{minipage}
\hfill
\begin{minipage}{0.48\textwidth}
\includegraphics[width=\textwidth]{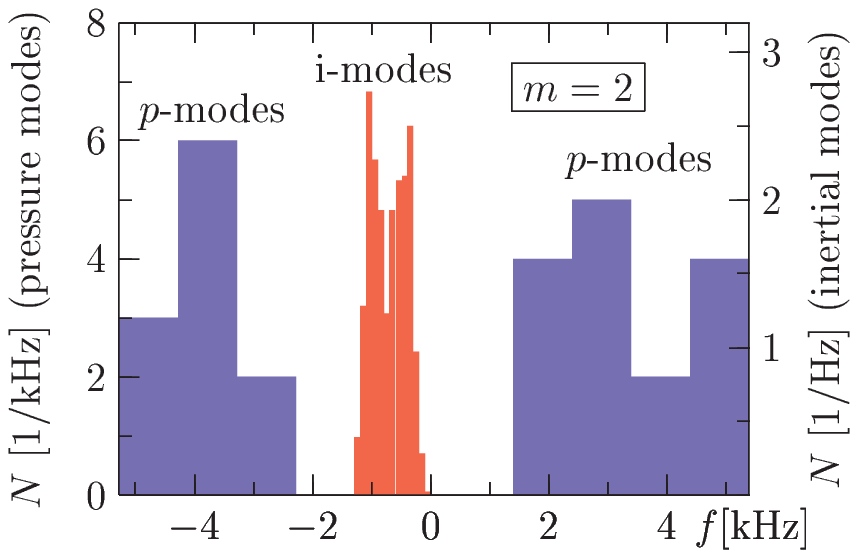}
\end{minipage}   \\[2.\baselineskip]
\begin{minipage}{0.48\textwidth}
\includegraphics[width=\textwidth]{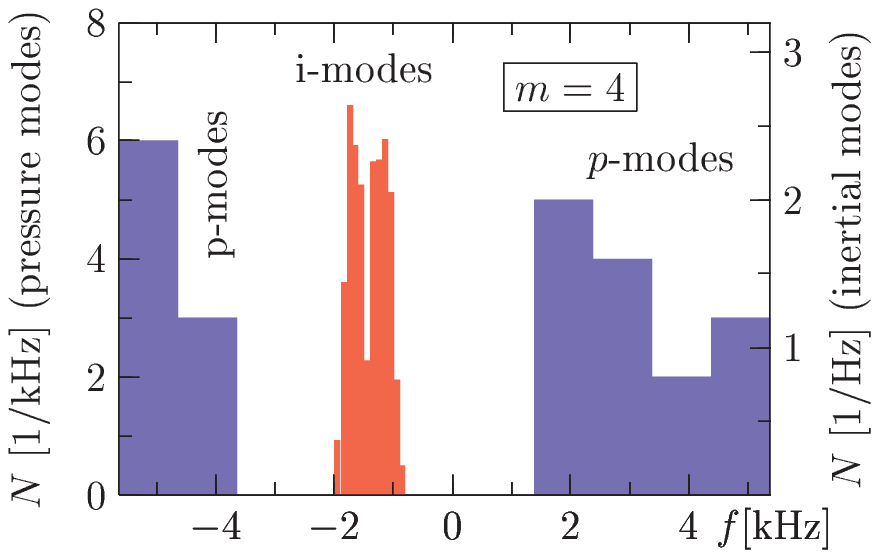}
\end{minipage}
\caption{\label{schema}
The histogram of both positive and negative solutions of the eigenvalue code
for $m=0,2$, and 4. 
The values for the inertial
modes (middle) are scaled differently for better presentation. 
}
\end{figure}

In the axisymmetric case a zero-frequency mode in the rotating frame
has 0 Hz for an inertial observer as well, corresponding to the mid-point
of the inertial mode frequency range.
For non-axisymmetric perturbations the center of the spectrum is expected to
shift to $-m\Omega$, which indeed appears to be the case in our results.
The corresponding inertial mode spectrum for a specific azimuthal index $m>0$
is however not completely symmetric since modes with different order $n$
or harmonic indices $\ell$ have different frequency changes.
This change is rather small for pressure modes.
As the azimuthal index $m$ can take arbitrarily large values,
the inertial modes can, according to the above, reach
arbitrarily large negative frequencies. Since these are equivalent to 
positive frequencies with a phase difference,
the frequency range of inertial modes will overlap with that of pressure modes.
An overview of the spectra for three different values of the azimuthal
index $m$ can be seen in Fig.\ \ref{schema}.

The oscillation frequencies are influenced not only by changes in the
equilibrium configuration of the star, but by other effects as well,
such as the frame dragging at the surface and the center
\citep{kojima}. These also scale linearly with $\Omega$.
Knowing the range of inertial mode frequencies as a function of the star's
rotational period in advance may be quite helpful for actual
observations of gravitational radiation emitted by these oscillations.
The frequencies of individual modes, such as the fundamental $r$-mode, are
not examined here, they will form the focus of a forthcoming paper.

\section{Conclusions}
\label{concl}
We have presented a code to numerically compute oscillation
frequencies of rapidly rotating relativistic stars for both axisymmetric
and non-axisymmetric perturbations, using the relativistic Cowling
approximation. 

For the polytropic equations of state that we have employed here, we
found an infinite set of pressure modes with a range of frequencies
bounded only from below at about 2kHz.
In addition, there is a presumably infinite set of solutions at lower
frequencies in a well defined range which corresponds to inertial modes.
In the axisymmetric case ($m=0$) this frequency range is symmetric
around zero. It extends to a maximum frequency which depends linearly on
the star's rotation rate $\nu_{star}$ and approaches $2\nu_{star}$ for
less relativistic stars.
For non-axisymmetric perturbations oscillation frequencies change towards
negative numbers. This affects pressure modes only slightly, while the
inertial modes now all have frequencies below zero.

It is not clear whether this low-frequency part of the spectrum is
discrete or continuous, or a combination of both. The dense and somewhat
uneven distribution of the eigenvalues is an indication that the actual
structure of the physical spectrum may be quite complicated.
Further investigation is
needed into the question how a spectrum with a continuous part can be
studied numerically. Furthermore, explicit time evolution of
perturbations of the same configurations could provide an independent
means of studying their spectrum and checking the results we have
presented here.
Identification of individual modes in this part of the spectrum will be
studied in a forthcoming paper.

Applying realistic equations of state requires only slight
modification of the code and will shift the frequencies to some extent,
but the overall picture should remain the same.  The effect of a
non-barotropic equation of state on the inertial mode spectrum presents
a very interesting question for further study.

Finally, for an accurate calculation of the oscillations frequencies,
the disposal of the Cowling approximation will be necessary. This is
important not just to avoid deviation of the calculated frequencies
introduced by the Cowling approximation, which we expect to be small,
but also to ensure the consistency of the problem (see
Section \ref{intro}). The major problem lies not in the perturbation
equations becoming more complex, but in defining the boundary conditions
in the spacetime outside the star. Observation of potentially
unstable modes with the now-operational gravitational-wave detectors is
possible only if precise understanding of these modes and their frequency
range is available as a basis for the analysis of the gravitational wave data.


\begin{acknowledgments}
We are grateful to K.\ Kokkotas for supporting this study, for inspiring
discussions and careful reading of the manuscript.
J.\ Frauendiener and N.\ Stergioulas provided key advice concerning
numerical procedures.
We also thank S.\ 'I.\ Yoshida, F.\ K.\ Lamb, and the anonymous referee for useful interactions.
This work was mainly supported by the german science foundation (DFG)
through the Transregio Sonderforschungsbereich 7, and in part by
NASA grant NAG~5-12030, NSF grant AST~0098399, 
and the funds of the Fortner Endowed Chair at the University of Illinois.
\end{acknowledgments}


\begin{widetext}
\appendix
\section{The full system of euations for the fluid perturbations}
\label{eqs}
\begin{eqnarray}
\partial_t F&=&\frac{imF}{Z\Pi-AB}\Biggl\{B\frac{\left[e^{2\rho}+r^2\sin^2{\theta}\omega(\Omega-\omega)\right]^2+e^{2\rho}\Omega^2r^2\sin^2{\theta}}{r^2\sin^2{\theta}e^{\gamma+\rho}u^0}
-(1+C_s^2)\Omega Z\left[e^{2\rho}+r^2\sin^2{\theta}\omega(\Omega-\omega)\right]\Biggr\}\nonumber\\
&&-\frac{imV}{Z\Pi-AB}\Biggl\{\Pi\frac{\left[e^{2\rho}+r^2\sin^2{\theta}\omega(\Omega-\omega)\right]^2+e^{2\rho}\Omega^2r^2\sin^2{\theta}}{r^2\sin^2{\theta}e^{\gamma+\rho}u^0}
-(1+C_s^2)\Omega A\left[e^{2\rho}+r^2\sin^2{\theta}\omega(\Omega-\omega)\right]\Biggr\}\nonumber\\
&&-C_s^2u^0\frac{e^{\gamma-\rho}}{e^{2\alpha}}\Biggl\{\frac{e^{2\rho}+r^2\sin^2{\theta}\omega(\Omega-\omega)}{(u^0)^2e^{\gamma-\rho}}\left(\frac{3}{2}\partial_r\gamma-\frac{1}{2}\partial_r\rho+\frac{3}{r}\right)-r^4\sin^4{\theta}\left(\Omega-\omega\right)^3\partial_r\omega\nonumber\\
&&-e^{2\rho}r^2\sin^2{\theta}\left[\omega\partial_r\omega+\Omega(\Omega-\omega)\left(\partial_r\rho-\frac{1}{r}\right)\right]+\frac{e^{2\rho}}{(u^0)^2e^{\gamma-\rho}}\left(\partial_r\rho-\frac{1}{r}\right)\Biggr\}f_1\nonumber\\
&&-\frac{C_s^2}{r^2}u^0\frac{e^{\gamma-\rho}}{e^{2\alpha}}\Biggl\{\frac{e^{2\rho}+r^2\sin^2{\theta}\omega(\Omega-\omega)}{(u^0)^2e^{\gamma-\rho}}\left(\frac{3}{2}\partial_{\theta}\gamma-\frac{1}{2}\partial_{\theta}\rho+2\cot{\theta}\right)-r^4\sin^4{\theta}\left(\Omega-\omega\right)^3\partial_{\theta}\omega+\nonumber\\
&&-e^{2\rho}r^2\sin^2{\theta}\left[\Omega(\Omega-\omega)\left(\cot{\theta}-\partial_{\theta}\rho\right)-\omega\partial_{\theta}\omega\right]+\frac{e^{2\rho}}{(u^0)^2e^{\gamma-\rho}}\left(\partial_{\theta}\rho-\cot{\theta}\right)\Biggr\}f_2\nonumber\\
&&-\frac{C^2_s}{u^0}e^{-2\alpha}\left[e^{2\rho}+r^2\sin^2{\theta}\omega\left(\Omega-\omega\right)\right]\left(\partial_r f_1+\frac{1}{r^2}\partial_{\theta}f_2\right)
\end{eqnarray}
\begin{eqnarray}
\partial_t V&=&-\frac{im}{Z\Pi-AB}\Biggl\{B\left(\frac{\omega}{(u^0)^2 e^{\gamma-\rho}}-2\Omega e^{2\rho}\right)+u^0e^{\gamma+\rho}Z\left[C_s^2e^{2\rho}+r^2\sin^2{\theta}(\Omega-\omega)(C_s^2\omega+\Omega)\right]\Biggr\}F\nonumber\\
&&+\frac{im}{Z\Pi-AB}\Biggl\{\Pi\left(\frac{\omega}{(u^0)^2 e^{\gamma-\rho}}-2\Omega e^{2\rho}\right)+u^0e^{\gamma+\rho}A\left[C_s^2e^{2\rho}+r^2\sin^2{\theta}(\Omega-\omega)(C_s^2\omega+\Omega)\right]\Biggr\}V\nonumber\\
&&-C_s^2r^2\sin^2{\theta}\frac{\Omega-\omega}{e^{2\alpha}}e^{\gamma+\rho}\left\{\frac{3}{2}\partial_r\gamma-\frac{1}{2}\partial_r\rho+\frac{3}{r}+(u^0)^2e^{\gamma+\rho}\left[-\partial_r\rho+\partial_r\ln{(\Omega-\omega)}+\frac{1}{r}\right]\right\}f_1\nonumber\\
&&-C_s^2\sin^2{\theta}\frac{\Omega-\omega}{e^{2\alpha}}e^{\gamma+\rho}\left\{\frac{3}{2}\partial_{\theta}\gamma-\frac{1}{2}\partial_{\theta}\rho+2\cot{\theta}+(u^0)^2e^{\gamma+\rho}\left[-\partial_{\theta}\rho+\partial_{\theta}\ln{(\Omega-\omega)}+\cot{\theta}\right]\right\}f_2\nonumber\\
&&-C_s^2r^2\sin^2{\theta}(\Omega-\omega)e^{\gamma+\rho}e^{-2\alpha}\left(\partial_rf_1+\frac{1}{r^2}\partial_{\theta}f_2\right)
\end{eqnarray}
\begin{eqnarray}
\partial_tf_1&=&\Biggl\{\frac{B}{C_s^2}\;\frac{2(\Omega-\omega)\left(\partial_r\rho-\frac{1}{r}\right)+e^{-2\rho}Z\partial_r\omega}{Z\Pi-AB}+\frac{\partial_r\ln{(p+\epsilon)}}{u^0(\Pi-AB/Z)}-\frac{1}{u^0}\partial_r\left(\frac{Z}{Z\Pi-AB}\right)\Biggr\}F\nonumber\\
&&-\Biggl\{\frac{\Pi}{C_s^2}\;\frac{2(\Omega-\omega)\left(\partial_r\rho-\frac{1}{r}\right)+e^{-2\rho}Z\partial_r\omega}{Z\Pi-AB}+\frac{\partial_r\ln{(p+\epsilon)}}{u^0(Z\Pi/A-B)}-\frac{1}{u^0}\partial_r\left(\frac{A}{Z\Pi-AB}\right)\Biggr\}V\nonumber\\
&&-\frac{\partial_rF}{u^0(\Pi-AB/Z)}+\frac{\partial_rV}{u^0(Z\Pi/A-B)}-im\Omega f_1
\end{eqnarray}
\begin{eqnarray}
\partial_tf_2&=&\Biggl\{\frac{B}{C_s^2}\;\frac{2(\Omega-\omega)\left(\partial_{\theta}\rho-\cot{\theta}\right)+e^{-2\rho}Z\partial_{\theta}\omega}{Z\Pi-AB}+\frac{\partial_{\theta}\ln{(p+\epsilon)}}{u^0(\Pi-AB/Z)}-\frac{1}{u^0}\partial_{\theta}\left(\frac{Z}{Z\Pi-AB}\right)\Biggr\}F\nonumber\\
&&-\Biggl\{\frac{\Pi}{C_s^2}\frac{2(\Omega-\omega)\left(\partial_{\theta}\rho-\cot{\theta}\right)+e^{-2\rho}Z\partial_{\theta}\omega}{Z\Pi-AB}+\frac{\partial_{\theta}\ln{(p+\epsilon)}}{u^0(Z\Pi/A-B)}-\frac{1}{u^0}\partial_{\theta}\left(\frac{A}{Z\Pi-B}\right)\Biggr\}V\nonumber\\
&&-\frac{\partial_{\theta}F}{u^0(\Pi-AB/Z)}+\frac{\partial_{\theta}V}{u^0(Z\Pi/A-B)}-im\Omega f_2
\end{eqnarray}
\end{widetext}

where \begin{eqnarray*}
A&=&\left(2\Omega e^{2\rho}-\frac{\omega}{(u^0)^2e^{\gamma-\rho}}\right)\frac{1}{e^{\gamma+\rho}u^0}\\
B&=&r^2\sin^2{\theta}(\Omega-\omega)e^{\gamma+\rho}u^0(1+C_s^2)\\
\Pi&=&e^{2\rho}+r^2\sin^2{\theta}(\Omega-\omega)\left(C_s^2\Omega+\omega\right)\\
Z&=&e^{2\rho}+r^2\sin^2{\theta}(\Omega-\omega)^2\\
\left[(u^0)^2 e^{\gamma-\rho}\right]^{-1}&=&e^{2\rho}-r^2\sin^2{\theta}(\Omega-\omega)^2
\end{eqnarray*}
and $\partial_r,\partial_{\theta}$ stand for the partial derivatives with respect to
the radial and angular coordinate.

\section{Discussion of some numerical difficulties}
\label{sec:numdisc}

The Cowling approximation does not include the damping of oscillations
by emission of gravitational waves, and no other damping mechanism is
included in our models. Therefore, all eigenfrequencies should be purely
real. However, the code we use is designed for calculating complex
frequencies. The imaginary parts of the computed values should therefore
turn out to be close to zero, on the order of machine accuracy
($\sim\leq 10^{-14}$). This is indeed the case for the majority of the
solutions we found. Some of the eigenvalues in the frequency range of
modes, however, have a non-negligible imaginary part of $\sim\leq 1\%$
of the real part.

We are thus facing the question if some error occurred in the
implementation of the perturbation equations into the code. The
perturbation equations where first computed using {\texttt Maple}. The
resulting expressions were then simplified by hand and implemented
numerically. As an alternative, we inserted the Maple output directly
into the code without any simplification. The spurious imaginary parts
of inertial mode frequencies then became larger, rather than smaller. We
are therefore confident that they are not the result of some actual
error in implementing the equation, but a consequence of truncation
error, especially in conjunction with a bad numerical condition of the
coefficient matrix $\mathcal{A}$.

Deviations from the expected convergence were found, while eigenfunctions
were also seen to be changing form in the rotating case. Testing our code
by relabeling the eigenvectors revealed the same results, which strengthens
its correctness.
Still, considerably more work is required to
clarify the question how numerical results can help to reveal the
complicated structure of the spectrum we are studying here.

\end{document}